\documentclass{article}
\usepackage{spconf,amsmath,graphicx}
\usepackage{multirow}
\usepackage{subfigure}
\usepackage{multicol}
\usepackage{array}
\usepackage{float}
\usepackage{hyperref}

\def\b1{{\mathbf 1}}

\title{Knowledge infused cascade convolutional neural network for segmenting retinal vessels in volumetric optical coherence tomography
}

\name{ Liyang Fang$^{1,4}$, Jianlong Yang$^{1^*}$, Lei Mou$^{1}$,  Huihong Zhang$^{1}$, Zhenjie Chai$^{1}$, Zhi Chen$^{2^*}$, Jiang Liu$^{1,3}$ \thanks{Thanks to Ningbo 3315 Innovation team grant for funding.}}

\address{ $^{1}$Cixi Institute of Biomedical Engineering, \\
Ningbo Institute of Industrial Technology, Chinese Academy of Sciences, China\\
$^{2}$Department of Ophthalmology, Fudan University Eye and ENT Hospital , China\\ 
$^{3}$ Southern Unerversity of Science and Technology, China\\
$^{4}$ School of Electrical Engineering, Southwest Jiaotong University , China}

\begin{document}

\maketitle

\begin{abstract}
We present a cascade deep neural network to segment retinal vessels in volumetric optical coherence tomography (OCT). Two types of knowledge are infused into the network for confining the searching regions. (1) Histology. The retinal vessels locate between the inner limiting membrane and the inner nuclear layer of human retina. (2) Imaging. The red blood cells inside the vessels scatter the OCT probe light forward and form projection shadows on the retinal pigment epithelium (RPE) layer, which is avascular thus perfect for localizing the retinal vessel in transverse plane. Qualitative and quantitative comparison results show that the proposed method outperforms the state-of-the-art deep learning and graph-based methods. This work demonstrates, instead of modifying the architectures of the deep networks, incorporating proper prior knowledge in the design of the image processing framework could be an efficient approach for handling such specific tasks.
\end{abstract}


\section{Introduction}
\label{Introduction}
Localization and segmentation of retinal vessels are critical for the diagnosis and research of ocular diseases, such as glaucoma, diabetic retinopathy, and age-related macular degeneration \cite{adhi2013optical}. The methods of automatic segmentation the retinal vessel from 2D retinal image techniques [e.g., fundus photography and scanning laser ophthalmoscopy (SLO) ]have been well developed \cite{fraz2012blood}, but the 2D fundus images can not provide the depth information, thus are unable to locate the vessel in specific retinal layers. 

\begin{figure}[h t p]
    \centering
    \includegraphics[width=3.2 in]{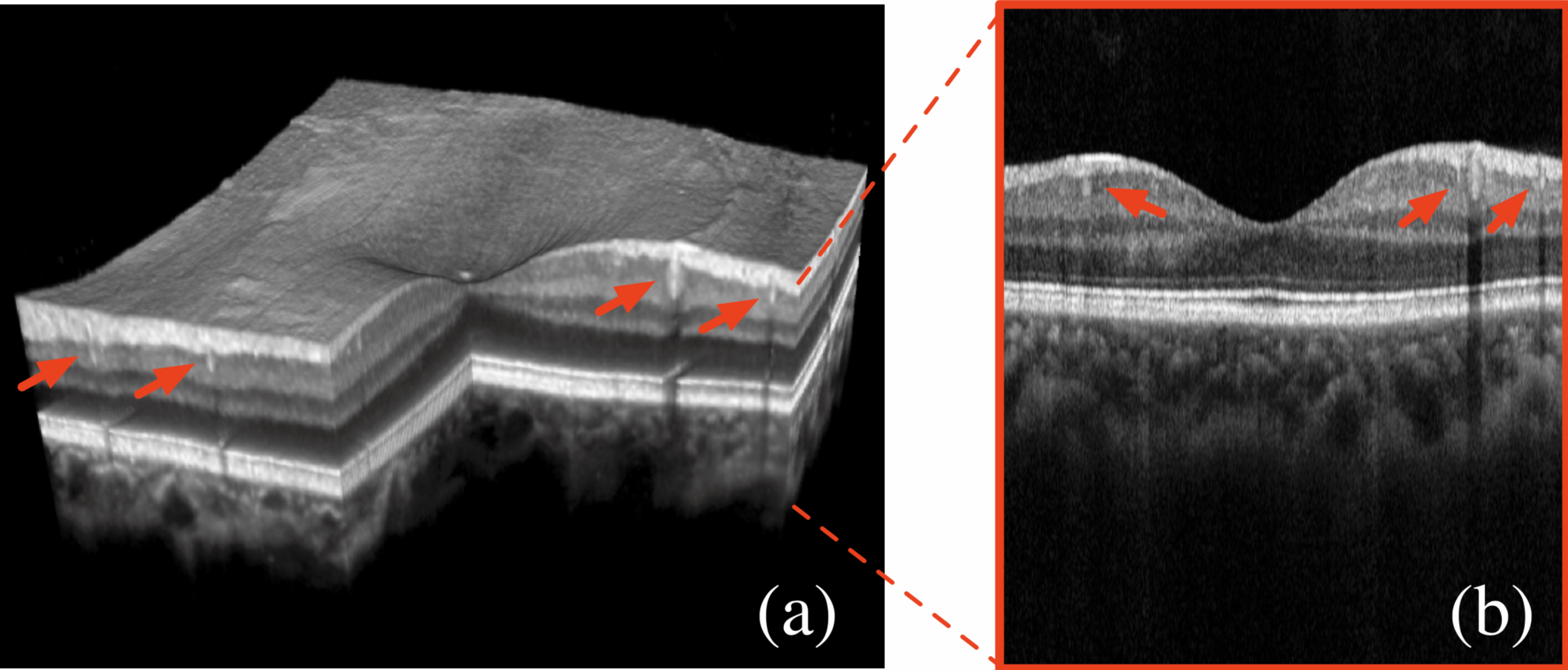}
    \caption{(a) OCT 3D volume. (b) OCT B-scan. The red arrows refer to the retinal vessels.}
    \label{fig:introduction}
\vspace{-0.4 cm}
\end{figure}

\indent Optical coherence tomography (OCT) is a non-invasive 3D imaging modality of the retina. However, the automatic segmentation of the retinal vessels in OCT is quite challenging.  Figure \ref{fig:introduction} demonstrates a OCT 3D volume (a) and a representative B-scan (b). The red arrows pointing to the positions of the retinal vessels which are small and point-like. 
The vessels have a limited contrast because of the reflectance of the surrounding tissue, which makes it difficult to perform automatic segmentation.
The emergence of OCT angiography (OCTA) partially eases the requirement of automatic vessel segmentation techniques at the expense of expensive and complex upgrades in OCT hardware and algorithms, but the sharply increased cost of the imaging system makes OCTA still unavailable in the majority of hospitals and eye clinics.\\
\begin{figure*}[ht!]
\vspace{-0.3cm}
    \centering
    \includegraphics[scale = 0.35]{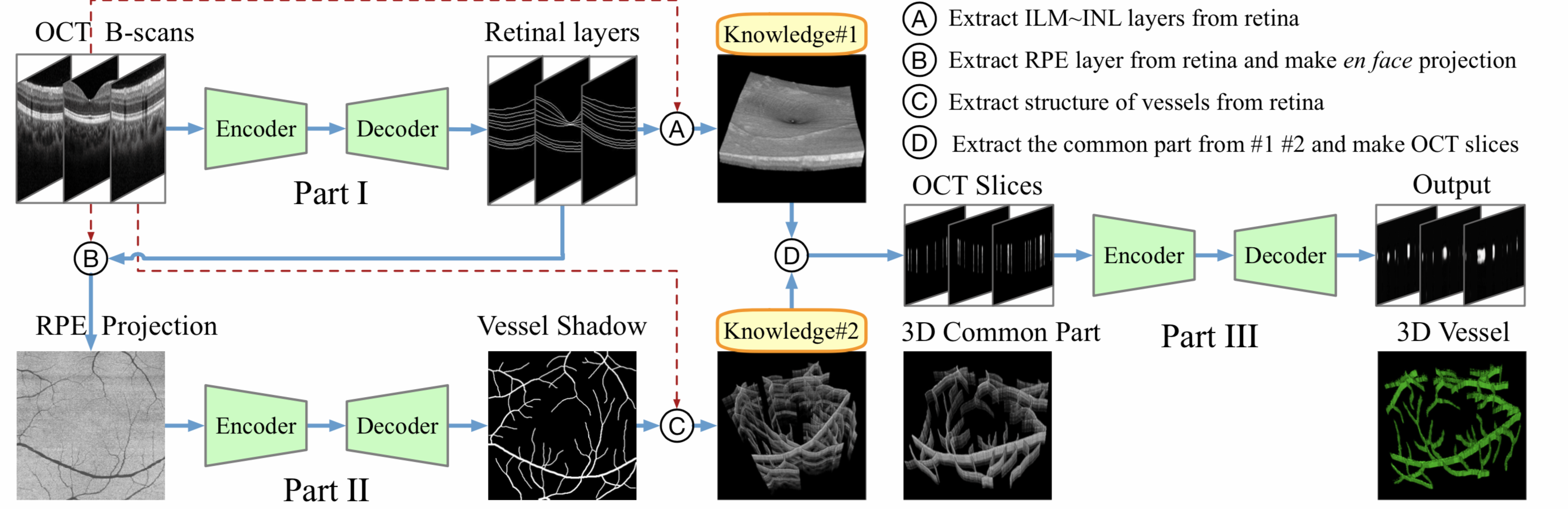}
    \caption{The overall architecture of the proposed method.} 
    \label{fig:overall}
\vspace{-0.3cm}
\end{figure*} 
\indent Different from the well-developed 2D vessel segmentation techniques, there are limited number of researches working on the 3D vessel segmentation of the OCT data. Hu \textit{et al.} proposed a trangular-mesh-based graph search method for segmenting 3D retinal vessels in the region near the neural canal opening \cite{hu2010automated}. Pilch \textit{et al.} approximated the lateral vessel positions by a shadowgraph and used active shape model to segment blood vessel contours in axial direction \cite{pilch2012automated}. After removing the outer retinal layers, the shadowgraph differentiates the vessel and non-vessel positions by intensity variations caused by the retinal shadows \cite{wehbe2007automatic}. Reif \textit{et al.} combined OCT and SLO to segmented the vessels based on brightness variation \cite{reif2014motion}. Also based on the shadowgraph, Guimaraes \textit{et al.} classified the A-lines into vessel and non-vessel using  support vector machine \cite{guimaraes2015three}.\\
\indent In recent years, deep learning has achieved massive successes in the segmentation and classification of medical images \cite{shen2017deep}. However, to the best of our knowledge, none of the deep learning techniques has been used in the vessel segmentation of retinal OCT volumes.\\
\indent In this paper, we propose a knowledge infused cascade deep network for segmenting the retinal vessels in volumetric OCT. Instead of modifying the architectures of the deep networks for improving the segmentation performance as demonstrated in the previous works \cite{long2015fully,badrinarayanan2017segnet, gu2019net}, we incorporate two types of knowledge priors in the segmentation network, which employ the U-shape convolutional network (U-Net) \cite{ronneberger2015u} as the backbone.  Combining the vessel's histology knowledge of the depth position and OCT imaging knowledge of the transverse position could significantly reduce the burden of searching, thus improving the segmentation performance.

\section{Method}
\label{Method}
Figure \ref{fig:overall}  illustrates the overall architecture of the proposed method, which primarily contains the segmentation of the retinal layers (Part \uppercase\expandafter{\romannumeral1}), the \textit{en face} projection of the RPE layer and the segmentation of the vessel shadows (Part \uppercase\expandafter{\romannumeral2}), and the infusion of the medical knowledge and the segmentation of the 3D vessels (Part \uppercase\expandafter{\romannumeral3}).

\subsection{U-Net segmentation modules} 
The segmentation tasks mentioned above were fulfilled by three U-Net modules, which were labeled and trained separately. The U-Net is an efficient fully convolutional network for medical image segmentation. It has an encoder--decoder architecture (as shown in Figure \ref{fig:overall} ) with four times downsampling and convolution for feature extraction and symmetrical upsampling and convolution to restore the feature map to the resolution of the original image. It further employs skip connections to recover the full spatial resolution at the network output \cite{drozdzal2016importance}.\\
\begin{figure}[h!]
\vspace{-0.4 cm}
    \centering
    \includegraphics[width=2.8in, height= 1.8in]{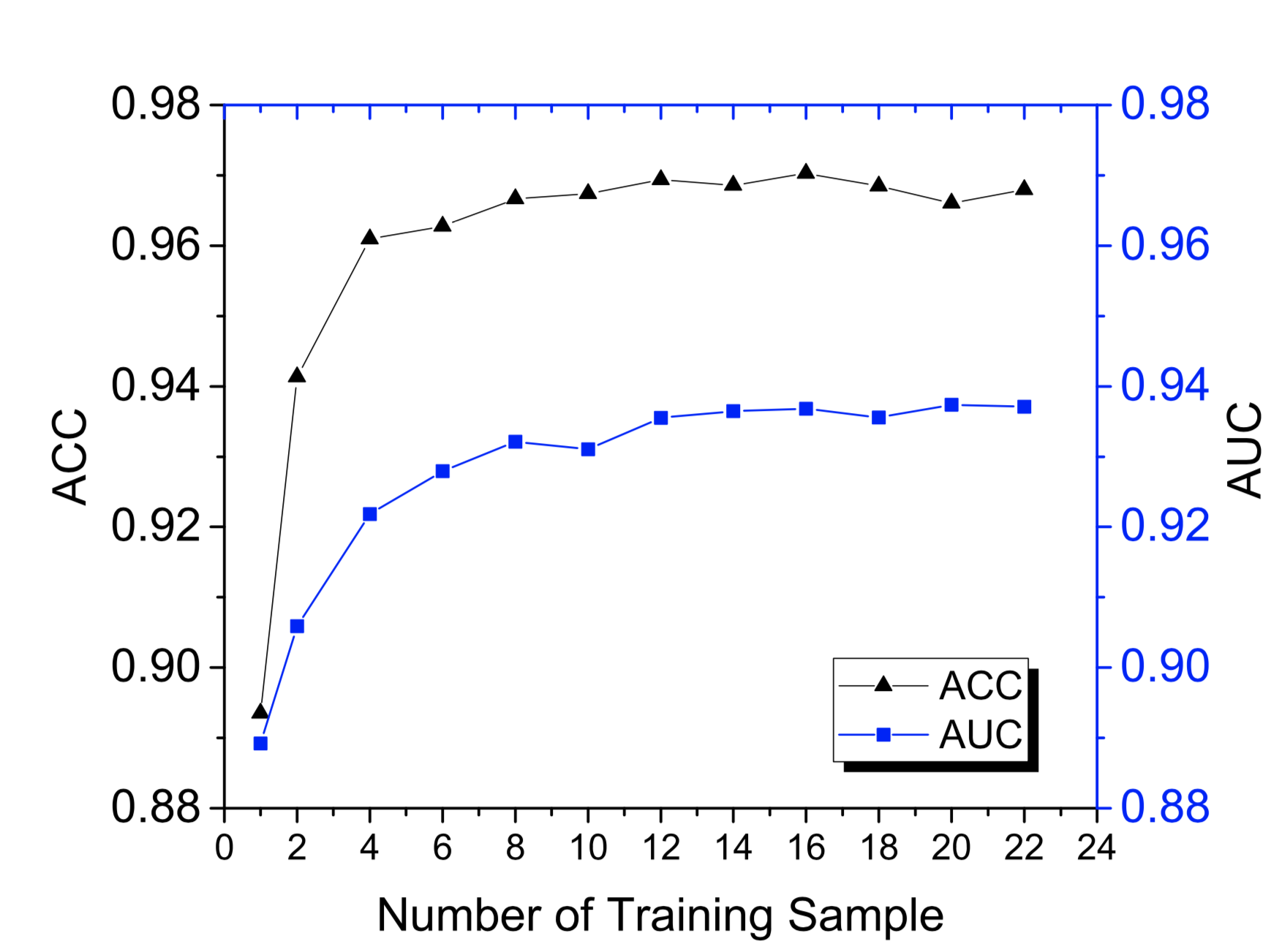}
    \caption{The ACC and AUC performance of the U-Net on the segmentation of the RPE vessel mask.}
    \label{fig:Graph2}
\vspace{-0.4 cm}
\end{figure}

\indent The pixel-level manual annotation for the segmentation tasks is tedious and time-consuming. The U-Net has demonstrated its capacity in achieving high segmentation accuracy by only using dozens of labeled images \cite{ronneberger2015u}. In this study, we further found a stable segmentation performance might be achieved by merely using a few images in the training. Figure \ref{fig:Graph2} shows the accuracy (ACC) and the area under the receiver operating characteristic curve (AUC) performance of the U-Net on the segmentation of the RPE vessel shadows. \\
\indent Through observations from Figure \ref{fig:Graph2}, we can conclude that ACC and AUC  trend to be stable when the number of training images is large than 5. We speculate this phenomenon is related to the high uniformity of the morphological patterns of the RPE vessel shadows among different OCT volumes. In comparison, the retinal layer and B-scan vessel segmentation tasks, which have significant variations of the morphological patterns, still need hundreds of images for the convergence of the training networks.
\begin{figure*}[ht!]
\vspace{-0.3 cm}
    \centering
    \includegraphics[width=6.7in, height=2.2in]{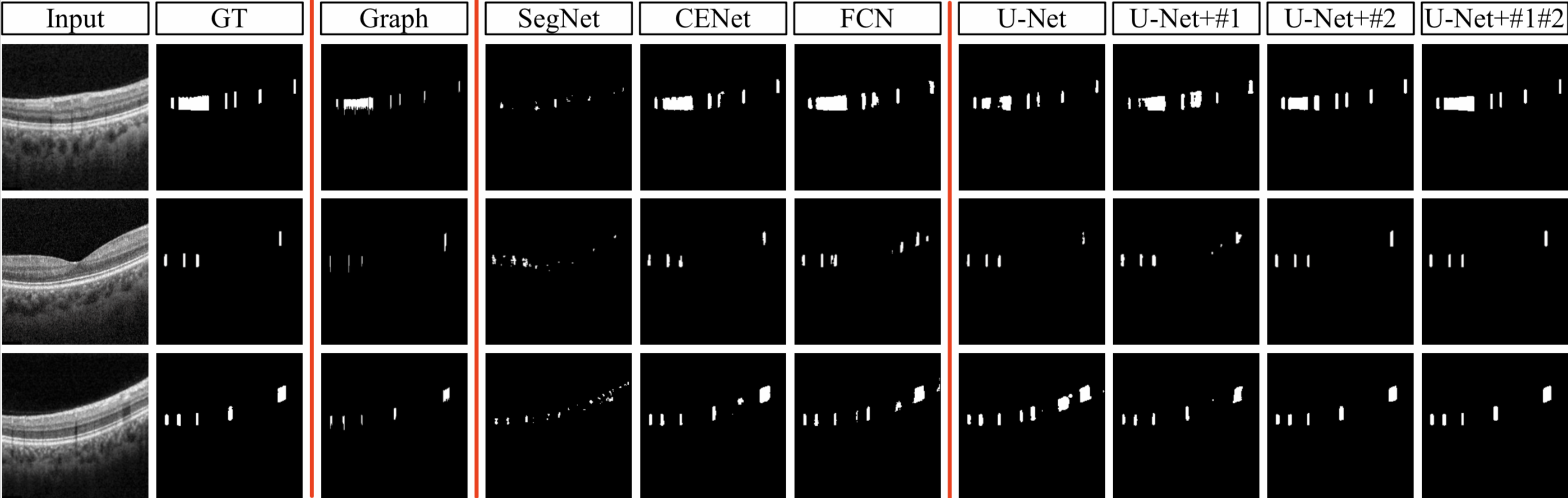}
    \caption{Qualitative comparison of the B-scan segmentation results. GT is the ground truth. \#1 and \#2 are the histology and imaging knowledge, respectively.}
    \label{fig:seg_result}
\vspace{-0.3 cm}
\end{figure*}

\subsection{Infusion of histology and imaging knowledge}
\indent Based on the histology of human retina, which were further confirmed by an \textit{in vivo} ocular circulation study using OCTA \cite{campbell2017detailed}, the vessels and capillaries locates at the outer retina, which is the histology knowledge (\#1) we infused into the proposed framework. We employ the segmented boundaries of the inner limiting membrane and the inner nuclear layer to form the longitudinal mask for searching the vessels. \\
\indent On the other hand, in the optics imaging of OCT, the anisotropy of the red blood cells inside the vessels cause strong forward scattering of the probe light, which result in shadow-like dark tails to the underneath layers from inner retina to choroid. Among the inner retinal layers, the RPE is the brightest layer that provide the best contrast for visualizing the shadows. This is the imaging knowledge (\#2) we infused into the proposed framework. Therefore, we employ the segmented boundaries of the RPE and Bruch's membrane for the 2D transverse mask of the vessel locations. 

\section{Experiments}
\subsection{Experiment setup}
\paragraph*{Datasets:} The dataset consists of 36 OCT volumes centered on from the both eyes of 18 healthy volunteers (age from 20 to 35; 13 males and 5 females). The human study protocol was approved by the Institutional Review Board of Cixi Institute of Biomedical Engineering, Chinese Academy of Sciences and followed the tenets of the Declaration of Helsinki. The OCT volumes were acquired by a Spectralis OCT system (Heidelberg Engineering GmbH). Each B-scan has $384 \times 496 $ pixels. The pixel-level segmentation ground truth images were manually annotated by two medical experts.
\vspace{-4mm}
\paragraph*{Evaluation metrics:} To facilitate objective performance evaluation of the proposed method, these quantitative metrics were calculated: intersection over union (IoU), AUC, sensitivity (SEN) = TP$/$(TP + FN), and accuracy(ACC) = (TP + TN)$/$(TP + FP + TN + FN), where TP, TN, FP, and FN are the true positive, true negative, false positive, and false negative, respectively. 

\paragraph*{Implementation:} The proposed method was implemented on PyTorch library with dual NVIDIA GPU (GeForce GTX 1080Ti). We employed the Adam as the optimizer \cite{kingma2014adam} with a weight decay 0.0005. The initial learning rate was set to 0.0001. We used poly learning rate policy where the learning rate is multiplied :
\begin{equation}
    L_r =  \left( 1- \frac{iter}{max\_iter} \right) ^ {power}
\end{equation}
 where \textit{iter} is number of iterations, \textit{max\_iter} is maximum number of iterations, \textit{power} is 0.9. All training images are rescaled to $368 \times 368$ pixels. We randomly selected 24 datasets for training and others for testing. 
 
\subsection{ Experimental Results}
To better justify the proposed method, we compared the segmentation results qualitatively and quantitatively using the proposed method with those using the state-of-the-art segmentation methods including the graph-based method (shadowgraph + brightness variation) \cite{guimaraes2015three} and the deep segmentation networks: FCN \cite{long2015fully}, segNet \cite{badrinarayanan2017segnet}, and CENet \cite{gu2019net}.
\vspace{-4mm}
\paragraph*{Qualitative comparison:} Figure \ref{fig:seg_result} demonstrates the B-scan vessel segmentation results. GT denotes the manually labeled ground truths. Each row in the Figure \ref{fig:seg_result} is a representative position of the OCT volume. Three red lines were used to separate the segmentation ground truth, the traditional graph-based method, the deep networks, and the proposed method, respectively. Through careful observations, we can see the graph-based method is capable of acquiring the transverse locations of the vessels precisely but unable to determine the longitudinal positions accurately. In case of the deep learning methods, the CENet and FCN can extract the longitudinal dimension of the vessel correctly but cause redundant segmentation region in the transverse dimension. In particular the SegNet is unable to correctly segment the vessels in both directions. The U-Net backbone of the proposed method has a similar performance as the CENet and FCN.
The proposed method performs best when infusing the histology (\#1) knowledge especially the imaging (\#2) knowledge into the framework.\\
\begin{figure}[h!]
\vspace{-0.3cm}
    \centering
    \includegraphics[width=3in, height=1.4in]{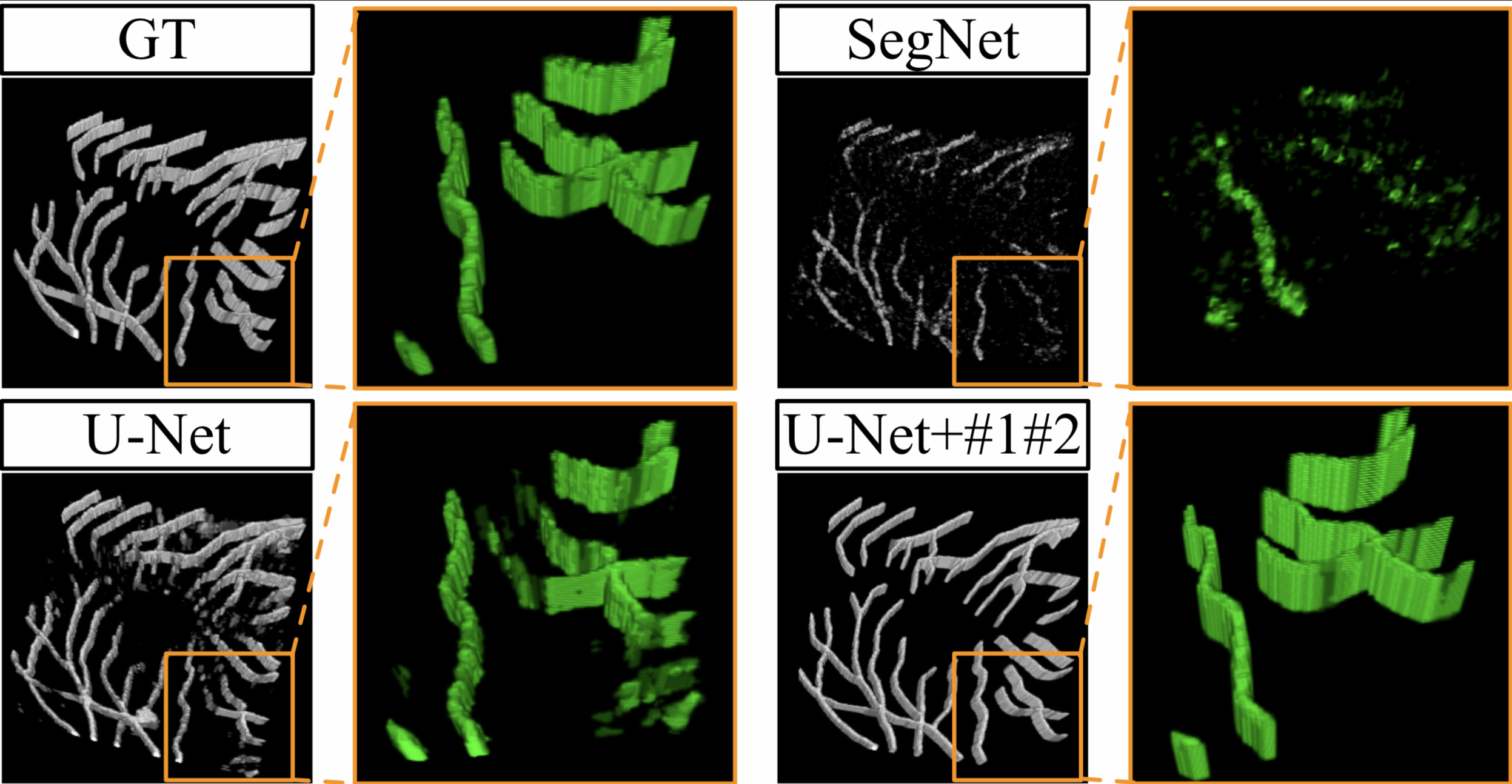}
    \caption{Qualitative comparison of the 3D segmentation results. GT is the ground truth. \#1 and \#2 are the histology and imaging knowledge, respectively.}
    \label{fig:result_3D}
\end{figure}

\indent We further demonstrate the 3D segmentation results in Figure \ref{fig:result_3D}. The orange boxes represent the zoom-in views of the 3D vasculature. It is obvious that  the proposed method achieves negligible discrepancy with the ground truth. While the results of the SegNet and U-Net have plenty of room for improvement.

\begin{table}[h t p]
\vspace{-0.3 cm}
\centering
\caption{Comparison of the segmentation methods.}\label{tab_octa}
\begin{tabular}{l | c c c c}
\hline
Methods & IoU & SEN & ACC & AUC \\
\hline
Graph-based ~\cite{guimaraes2015three} & 0.6473 & 0.8564 & 0.9965 &  -\\
\hline
SegNet ~\cite{badrinarayanan2017segnet} & 0.4234 & 0.5426 & 0.9965 &  0.9754\\
CENet ~\cite{gu2019net}& 0.5838 & 0.6942 & 0.9939 & 0.9759 \\
FCN ~\cite{long2015fully}& 0.6059 & 0.7101 & 0.9943 &  0.9747\\
\hline
U-Net ~\cite{ronneberger2015u} & 0.6149 & 0.7478 & 0.9942 &  0.9775\\
U-Net~+~\#1 & 0.6399 & 0.7543 & 0.9952 &  0.9793\\
U-Net~+~\#2 & 0.8851 & 0.9546 & 0.9979 &  0.9819\\
{\bfseries U-Net~+~\#1 + \#2} &{\bfseries 0.8971}  & {\bfseries 0.9662} & {\bfseries 0.9985} & {\bfseries 0.9821} \\
\hline
\end{tabular}
\end{table}
\vspace{-4mm}
\paragraph*{Quantitative comparison:} Table \ref{tab_octa} is the quantitative comparison of different segmentation methods using the metrics IoU, SEN, ACC, and AUC. For the IoU and SEN, their values are in accordance with the qualitative observation in Figure \ref{fig:seg_result} and Figure \ref{fig:result_3D} The graph-based method is better than the deep segmentation networks. The proposed method achieves a better performance with the incorporation of the prior knowledge.\\
\indent On the other hand, the small proportion of blood vessels in the OCT B-scans leads to an excessive TN cases in this binary segmentation task. So we can achieve very high ACC and AUC values for all the involved methods as shown in the table. Even though, the proposed method still outperforms other methods on these two metrics. \\

\paragraph*{Ablation study:} We did the ablation study by combining the two types of knowledge separately with the U-Net backbone. As shown in Figure \ref{fig:seg_result},  Figure \ref{fig:result_3D}, and Table 1, the infusion of the imaging knowledge (\#2), coarsely locating the vessels in the transverse plane, contributes more significantly to the accuracy of the vessel segmentation than the histology knowledge (\#1). This observation is in accordance with the previous works that strove for acquiring the transverse vessel information from other imaging modalities or the shadowgraph \cite{pilch2012automated,wehbe2007automatic,reif2014motion}.

\section{CONCLUSION}
\label{conclusion}
We have proposed a knowledge infused cascade network for extracting the retinal vessels from volumetric OCT. We have demonstrated that the U-Net as the backbone of the segmentation tasks, could achieve excellent accuracy with very few labeled data for training. The histology and imaging knowledge priors of the retinal vessels in OCT imaging have proven to be able to significantly benefit the accuracy of this task compared with the state of the art segmentation deep networks. This work has demonstrated that combining the deep networks with proper prior knowledge could be efficient in specific medical image processing tasks.

\bibliographystyle{IEEEbib}
\bibliography{Template_ISBI2018}

\end{document}